\documentclass[12pt]{article}
\usepackage{graphicx,amssymb, amsmath}
\newcommand{\mathsym}[1]{{}}

\input epsf.tex

\let\badcite=\cite
\def\cite{~\badcite}

\def\slashchar#1{\setbox0=\hbox{$#1$}           
   \dimen0=\wd0                                 
   \setbox1=\hbox{/} \dimen1=\wd1               
   \ifdim\dimen0>\dimen1                        
      \rlap{\hbox to \dimen0{\hfil/\hfil}}      
      #1                                        
   \else                                        
      \rlap{\hbox to \dimen1{\hfil$#1$\hfil}}   
      /                                         
   \fi}
    \def\slashword#1{\setbox0=\hbox{$#1$}        
  \dimen0=\wd0                                   
   \setbox1=\hbox{/} \dimen1=\wd1                
   \ifdim\dimen0>\dimen1                         
      \rlap{\hbox to \dimen0{\hfil\bf---\hfil}} %
      #1                                         %
   \else                                         
      \rlap{\hbox to \dimen1{\hfil$#1$\hfil}}    
      /                                          
    \fi}                                         %

\catcode`@=11
\newdimen\vbigd@men                             

\def\vbig#1#2{{\vbigd@men=#2\divide\vbigd@men by 2%
   \hbox{$\left#1\vbox to \vbigd@men{}\right.\n@space$}}}
\catcode`@=12

\catcode`@=11
\def\citenum#1{\csname b@#1\endcsname}
\catcode`@=12

\def\dofig#1#2{\centerline{\epsfxsize=#1\epsfbox{#2}}}
\begin{document}
\begin{titlepage}

\begin{flushright}
{SCUPHY-TH-08002}\\
\end{flushright}

\bigskip\bigskip

\begin{center}{\Large\bf\boldmath
Hidden Thresholds: A Technique for Reconstructing New Physics Masses at Hadron Colliders}
\end{center}
\bigskip
\centerline{\bf P. Huang, N. Kersting\footnote{Email: nkersting@scu.edu.cn}, H.H. Yang }
\centerline{{\it Physics Department, Sichuan University (Chengdu), P.R. China 610065}}

\bigskip

\begin{abstract}

We present an improved method of reconstructing New Physics (NP) masses from invariant mass endpoints.
While the traditional method focuses on a single NP decay, our method considers the
decays of two or more NP particles ($ABC...$) in a grander decay chain:
$ anything \to ABC... \to ... \to jets + leptons$. Though the center-of-mass energy $E_{CM}$ of `anything' varies unpredictably at a hadron collider, a sample of many events nonetheless expresses features of threshold production $E_{CM} = m_A + m_B + ...$: invariant masses constructed from the final jet and lepton momenta are correlated in a way that makes their threshold endpoints visually obvious in a scatterplot.
We illustrate this technique for the production of two neutralinos in the MSSM:
$anything \to \widetilde{\chi}_{i}^0 \widetilde{\chi}_{j}^0$ ($i,j=2,3,4$) which
subsequently decay via on- or off-shell sleptons to four leptons.
Assuming the relevant SUSY spectrum is below $1~TeV$ and squarks/gluinos eventually decay to neutralinos,
 our MC study shows that one low-luminosity year at the LHC ($10-30 fb^{-1}$) can quantitatively determine on- versus off-shell decays and find the relevant neutralino and slepton masses to less than 10 percent.

\end{abstract}

\newpage
\pagestyle{empty}

\end{titlepage}


\section{Introduction}

At a hadron collider such as the Tevatron at Fermilab or the Large Hadron Collider (LHC) at CERN, searches for New Physics (NP) states beyond the Standard Model (SM) must take into account the fact that partonic center of mass energies at these machines are not tunable (as they will be at a much anticipated $e^+ e^-$ linear collider\cite{linearcollide}) but vary continuously in principle from zero to the combined hadronic energies of approximately $2\, \hbox{TeV}$ and $14\, \hbox{TeV}$, respectively. Moreover,  many NP models predict the production of a long-lived particle that is likely to escape the detectors, carrying away missing energy. The traditional method of seeing a NP mass as a sharp resonance in a cross section is therefore not applicable, and we must consider other approaches to precision measurement of NP masses crucial for testing properties of an underlying fundamental theory.

One well-studied avenue is to construct invariant mass distributions of final jet or leptonic momenta in exclusive channels and study their endpoints. Even if some NP particles carry away missing energy in each event, endpoints of said distributions can be measured and matched to analytical functions of NP masses\cite{inv1}. There are several caveats to this method however. First, the exclusive channel under study must somehow be identified or assumed. Second, backgrounds must not interfere with endpoint measurement. Third, there may be some model-dependence in the method of fitting the endpoint on a 1-dimensional histogram. The first caveat is the most difficult to deal with, especially in a model such as the Minimal Supersymmetric SM (MSSM), where the gluino and squarks decay via literally hundreds of possible decay chains("cascades"). Studies of cascades which have enough endpoints to solve for MSSM masses, e.g.
  $\tilde{g} \to \tilde{q} q \to \widetilde{\chi}_{2}^0 q q \to
 \tilde{l}^\pm {l}^\mp q q \to l^\pm {l}^\mp q q \widetilde{\chi}_{1}^0$,  usually just focus on a few 'benchmark scenarios'\cite{benches,inv2} which, however, may not be what Nature chooses.
As for eliminating SM backgrounds, requiring a suitable number of hard jets and isolated leptons may suffice; NP backgrounds are more challenging and, if these can be reduced, typically also inflict damage in the region of the desired endpoint where rates are already low\footnote{One notable exception is the decay $A \to B~ l^\pm \to C ~l^\pm l^\mp$, where $A,B,C$ are NP particles, e.g. in the MSSM $\widetilde{\chi}_{2}^0, \tilde{l}, \widetilde{\chi}_{1}^0$. Here the dilepton invariant mass distribution \emph{rises} up to its endpoint and is therefore robust in the presence of diffuse backgrounds.}. This then brings up the third problem of how to fit the endpoint. Linear or Gaussian fits are most convenient, but very detailed study of cuts and detector effects are required to understand their general accuracy\cite{cuts}.

In this work we wish to show that there are two important features of NP particle production which, when used together, can greatly boost the efficacy of the endpoint method. First, if NP particles carry a new conserved charge, such as R-parity in the MSSM, they will be multiply-produced. We may, for example, consider inclusive decay chains of the form $\mathbb{X} \to A B \to \mathrm{jets} + \mathrm{leptons} $, where $A$ and $B$ are NP states and $\mathbb{X}$ is any system of particles with a sufficiently large total invariant mass:
 $m_\mathbb{X} \ge m_A + m_B$ (the case where  $m_\mathbb{X} = m_A + m_B$ is especially important and we designate this `threshold production.'). Since $\mathbb{X}$ usually includes at least one multi-particle system, $m_\mathbb{X}$ really is continuously-valued from the threshold value all the way up to the machine energy.
 Second, depending on $m_\mathbb{X}$ and the exact way in which $AB$ decay to the specified endstate,  invariant masses constructed from the final jet and lepton momenta attain endpoint values for certain kinematic configurations only.
  At threshold production, in particular, one special configuration will simultaneously maximize several invariant masses; collecting a large number of threshold decays, a 2-d or 3-d scatter plot of these invariants would exhibit a clustering around this `threshold point.' Yet threshold production is clearly only an infinitesimal possibility and superimposing events for `above-threshold' production ($m_\mathbb{X} > m_A + m_B$) would probably hide the threshold point (hereafter called a `hidden threshold'). Contrary to this intuition, however, we find that, for some invariant mass combinations, the hidden threshold is highly visible, being in fact fortified by above-threshold events. This
  allows us to directly measure\footnote{There is also no need to `fit' these endpoints, as we will see later.} values of threshold invariant mass endpoints which, as usual, can be used to constrain NP masses. Moreover, backgrounds should not be a problem since these have different invariant-mass correlations.

The precise mechanism of this Hidden Threshold (HT) technique is best illustrated by example, which in this work we
take to be
\begin{equation} \label{zizjdecay}
\mathbb{X}\to \widetilde{\chi}_{i}^0 \widetilde{\chi}_{j}^0(\to
    \tilde{\ell}^\pm {\ell}^\mp \to \ell^+ \ell^- \widetilde{\chi}_{1}^0) ~~~~~~~~(i,j=2,3,4)
\end{equation}
 i.e. neutralino pair-production (via any parent channel) and decay to leptons ($\ell = e,\mu$).
Section \ref{sec:invmass}  explains the basic HT theory in this case, followed
in Section \ref{sec:mc} by application  to Monte Carlo (MC) generated data simulating a low-luminosity($10-30~fb^{-1}$) run at the LHC for two different MSSM parameter points: one with on-shell slepton decays, and another with off-shell decays. Finally, Section \ref{sec:conc} summarizes these results and suggests many avenues for further application.

\section{The Hidden Threshold Technique} \label{sec:invmass}
\subsection{Base Case: Pure Higgs Decay}
Let us explain the HT technique in three steps, the first of which will be the case where $\mathbb{X}$ in
(\ref{zizjdecay}) is a single particle which therefore has a fixed mass, say a heavy Higgs boson ($H$ or $A$, hereafter designated 'Higgs'):
\begin{equation} \label{hdecay}
pp \to H/A \to \widetilde{\chi}_{i}^0 \widetilde{\chi}_{j}^0 \to
    \tilde{e}^\pm {e}^\mp \tilde{\mu}^\pm {\mu}^\mp
    \to e^+ e^- \mu^+ \mu^- \widetilde{\chi}_{1}^0 \widetilde{\chi}_{1}^0
\end{equation}
 For now, let us take the intermediate sleptons to be on-shell (see Sec.~\ref{subsec:off} below for the case of off-shell decays). We studied this decay chain in \cite{4l-inv}, and for the convenience of the reader we briefly recapitulate. From the four observable final lepton momenta in (\ref{hdecay}), one can define seven independent invariant mass combinations. These we defined as follows (leptons labelled as $1,1',2,2'$, see Fig.~\ref{fig:kin}):
 \begin{eqnarray}\label{avinv1}
  M_{4l}^2 &\equiv& (p_1 + p_{1'}+ p_2 + p_{2'})^2 \\
  \overline{M}_{{2l2l}}^4 &\equiv& \{(p_1 + p_{1'}- p_2 - p_{2'})^4 +
  (p_1 + p_{2'}- p_2 - p_{1'})^4 \\ \nonumber
  &&
  + (p_1 + p_{2}- p_{1'} - p_{2'})^4\} /3
     \\
  \overline{M}_{{l3l}}^4 &\equiv&  \{
  (p_1 + p_{1'}+ p_2 - p_{2'})^4 +
  (p_1 + p_{1'}+ p_{2'} - p_2)^4   \\ \nonumber
  &&
  + (p_1 + p_{2}+ p_{2'} - p_{1'})^4 +
  (p_{2}+ p_{2'}+ p_{1'} - p_{1})^4
 \} /4\\
  \overline{M}_{{l2l}}^4 &\equiv&  \{
  (p_1 + p_{1'}- p_2)^4 +
  (p_1 + p_{1'}- p_{2'} )^4
  + (p_1 + p_{2}- p_{2'})^4  \\ \nonumber
  &&
  +
  (p_{2}+ p_{2'}- p_{1'} )^4  +
  (p_1 - p_{1'}+ p_2)^4 +
  (p_1 - p_{1'}+ p_{2'} )^4  \\ \nonumber
  &&
  + (p_1 - p_{2}+ p_{2'})^4 +
  (p_{2}- p_{2'}+ p_{1'} )^4 +
  (p_{1'}+ p_2-p_1)^4  \\  \nonumber
 &&
  +(p_{1'}+ p_{2'}-p_1 )^4
  + (p_{2}+ p_{2'}-p_1)^4 +
  (p_{2'}+ p_{1'} -p_{2})^4  \}/12
    \\
  \overline{M}_{{3l}}^4 &\equiv&  \{
  (p_1 + p_{1'}+ p_2)^4 +
  (p_1 + p_{1'}+ p_{2'} )^4 \\ \nonumber &&
  + (p_1 + p_{2}+ p_{2'})^4 +
  (p_{2}+ p_{2'}+ p_{1'} )^4
 \}/4 \\
  \overline{M}_{{ll}}^4 &\equiv&
   \{
  (p_1 + p_{1'})^4 +
  (p_1 + p_{2'})^4 + (p_1 + p_{2})^4
   \\ \nonumber &&
  + ( p_2 + p_{2'})^4  + ( p_2 + p_{1'})^4 + ( p_{1'} + p_{2'})^4 \} /6
  \\
      a_4 & \equiv &  p_{1}^\mu p_{1'}^\nu p_{2}^\rho p_{2'}^\sigma \epsilon_{\mu \nu \rho \sigma} \label{avinv2}
\end{eqnarray}

  Adding the usual dilepton invariants $M_{ee}$ and $M_{\mu \mu}$ to this list, we studied all distributions (1-d histograms) and derived analytic formulae for their associated  endpoints (see \cite{4l-inv} for exact expressions, too lengthy to reproduce here). In MC simulations of LHC data (assuming a luminosity of $L= 300\, \hbox{fb}^{-1}$)  at several MSSM points where both neutralinos and both sleptons were degenerate,  endpoint precisions were not high enough to give better than 30\% determination of the unknown masses  $m_H,~  m_{\tilde{l}},~m_{{\tilde\chi_j}^0},~ {\rm and}~ m_{{\widetilde\chi_1}^0}$ (hereafter we abbreviate
  $m_s \equiv m_{\tilde{l}}$, $m_i \equiv m_{{\widetilde\chi_i}^0}$).
 This rather unexpectedly poor resolution from a seemingly over-constrained system(seven\footnote{We did not derive an endpoint formula for $a_4$; however, we could numerically see it did not make a difference.} constraints on four unknowns)
 owes to the highly non-linear form of the endpoint formulae, which generically yield a discrete set of solutions --- adding in smearing and detector effects melds this discrete set into the large continuous range quoted.
  However, if one of the masses were already known to 5\% accuracy, the degeneracy of the solutions breaks and the other three masses are likewise well-determined. Though our choice of invariants in (\ref{avinv1})-(\ref{avinv2}) is somewhat more complicated (yet more systematic) than is customary,
  the above is nonetheless a standard illustration of the usual invariant mass endpoint method.

\subsection{Threshold Higgs}

\begin{figure}[!htb]
\begin{center}
\dofig{4.10in}{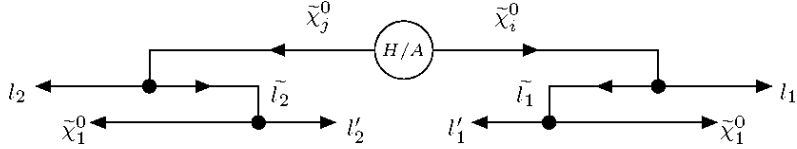}
\vskip -15.7cm
\end{center}
\caption{\small \emph{ Kinematic configuration which simultaneously maximizes  $M_{4l}$, $\overline{M}_{2l2l}$,
 $\overline{M}_{l2l}$, $\overline{M}_{3l}$, $\overline{M}_{ll}$, and $(M_{ee} \times M_{\mu\mu})$
  at threshold ($m_H = m_i + m_j$).}
}
 \label{fig:kin}
\end{figure}

Next, we consider what happens if the two neutralinos in (\ref{hdecay}) are produced at threshold\footnote{This situation was far from realized in \cite{4l-inv}, where the Higgs mass was set fairly high, $m_H \sim 400 - 600 ~\hbox{GeV}$.}, that
is when $m_H = m_i + m_j$ exactly. Now, it turns out, five of the invariants defined above ($M_{4l}$, $\overline{M}_{2l2l}$, $\overline{M}_{l2l}$, $\overline{M}_{3l}$, and $\overline{M}_{ll}$), in addition to
$M_{ee}$ and $M_{\mu\mu}$, are simultaneously maximal when the kinematic configuration shown in Fig.~\ref{fig:kin}
is realized.

\begin{figure}[!htb]
\begin{center}
\includegraphics[width=5.0in]{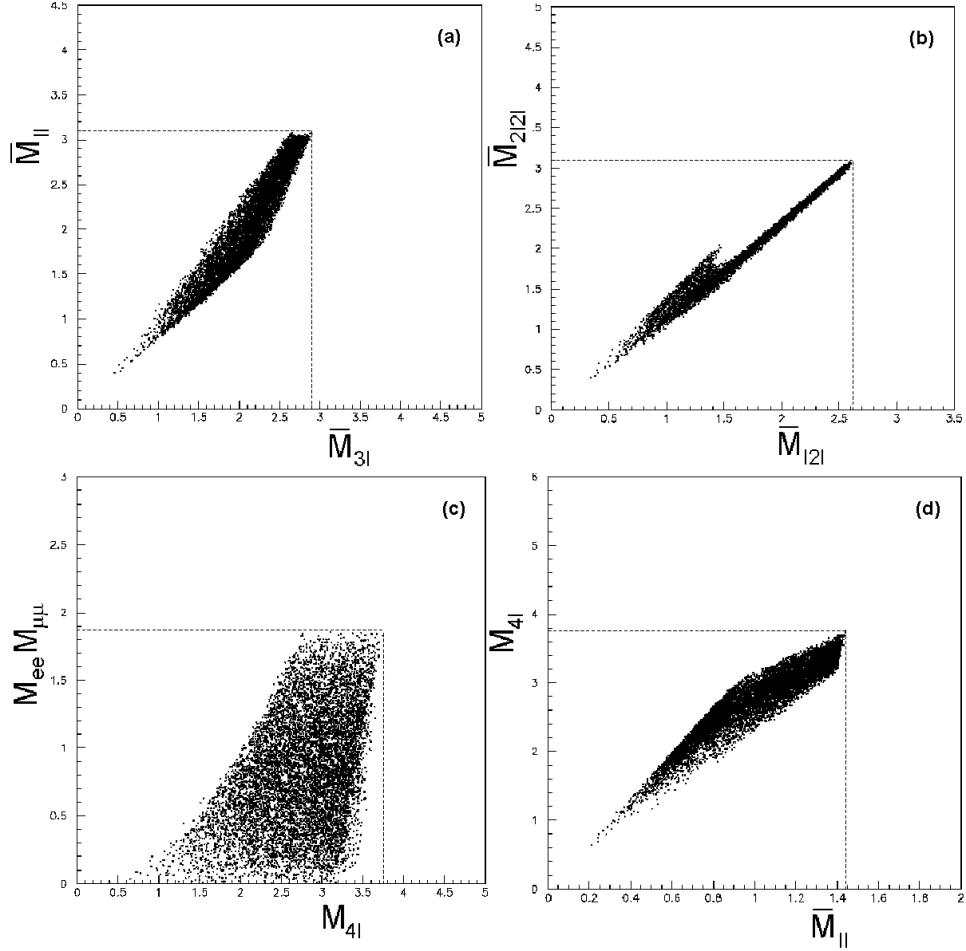}
\end{center}
\vskip -9.0cm
\caption{\small \emph{ Various correlations of four-lepton invariant masses at threshold ($m_H =  m_i + m_j$). This simulation is purely from relativistic kinematics with
$(m_1, ~m_i, ~m_j, ~m_s) ~=~ (1,~3,~4,~1.1)$ (units arbitrary). Dashed lines mark the maximum value of each invariant.}
}
 \label{fig:theory1}
\end{figure}

With a suitably large sample of events, we need only plot one invariant against another to get its maximal value, as shown in Figure \ref{fig:theory1}.  Note that the endpoints tend to lie at a fairly triangular apex, therefore well-approximated as the intersection of two tangents to the `correlation shape'. Moreover, this correlation retains its shape in the presence of  backgrounds, since these latter do not have the correct correlations among the various invariants and would tend to form a diffuse halo around the more concentrated signal shape. Compared to the traditional one-dimensional histogram approach, where backgrounds are harder to subtract and endpoints must be fit with a more arbitrary function, the reader can begin to appreciate that a technique using correlated invariants is much more powerful.
Though this case of threshold production is not particularly likely in the MSSM, it prepares us for the next step.

\subsection{Continuous Superposition of Higgses}

\begin{figure}[!htb]
\begin{center}
\includegraphics[width=4.5in]{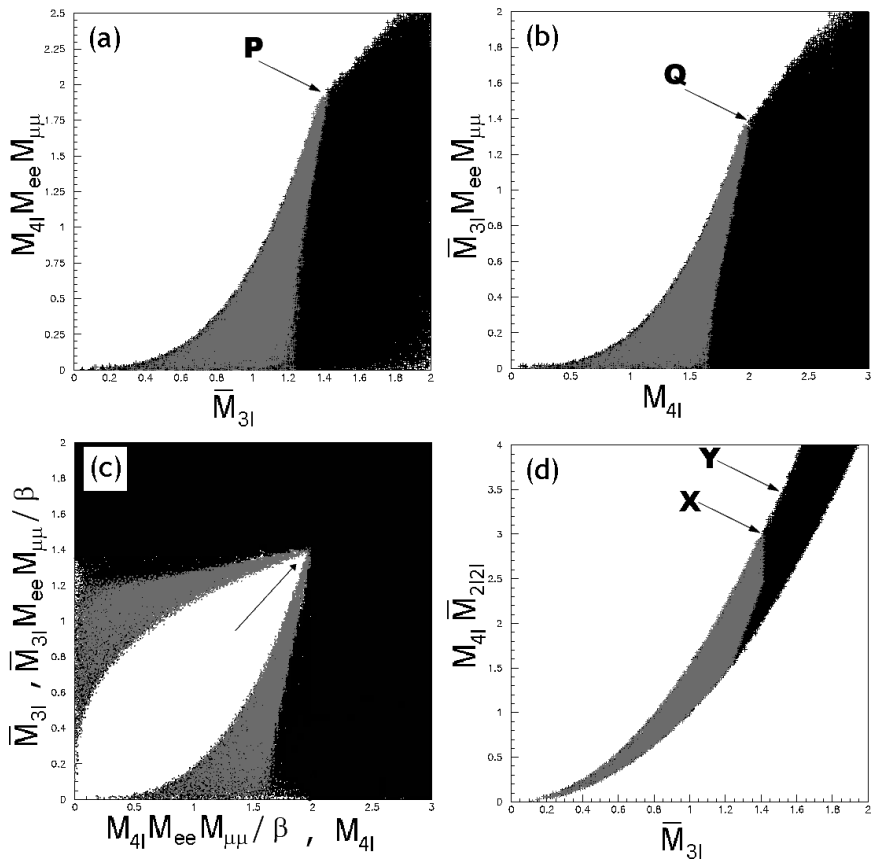}
\end{center}
\vskip -4.0cm
\caption{\small \emph{ Correlations of various invariant masses in a superposition of many Higgs decays (\ref{hdecay}) with a continuum of masses in the range
$ m_i + m_j  \le m_H < \infty$ (again from relativistic kinematics only, with
$(m_1, ~m_i=m_j, ~m_s) ~=~ (1,~2,~1.5)$ in arbitrary units). Points P and Q in (a) and (b) identify threshold endpoints; these can be found precisely with a `cat-eye' plot in (c), where we have abbreviated $\beta \equiv (M_{e e}^{max})\times(M_{\mu \mu}^{max})$. In (d) we show another useful correlation of invariants which not only gives threshold endpoints at X, but also `pseudo-threshold' endpoints at any point Y higher up on the envelope.
 }
}
 \label{fig:theory2}
\end{figure}

Consider now the case where not a single Higgs decays in (\ref{hdecay}), but rather a continuum of Higgs
with masses in the range $ m_i + m_j  \le m_H < \infty$. Fig.~\ref{fig:theory2} shows what to expect for several strategically-chosen\footnote{For the specific decay topology (\ref{zizjdecay}), it turns out that only certain correlations between invariants are easily analyzed, these being $M_{4l}~vs.~\overline{M}_{3l}M_{e e}M_{\mu \mu}$,  $\overline{M}_{3l}~vs.~{M}_{4l}M_{e e}M_{\mu \mu}$, $\overline{M}_{3l}~vs.~{M}_{4l} \overline{M}_{2l2l}$, $\overline{M}_{3l}~vs.~{M}_{4l} \overline{M}_{l2l}$, and $\overline{M}_{ll}~vs.~\overline{M}_{2l2l} \overline{M}_{l2l}$. } choices of invariants: threshold decays are plotted in gray, while all above-threshold decays are plotted in black\footnote{Threshold decays obviously contribute only infinitesimally to the total shape, but for the
sake of seeing how these are distributed compared to above-threshold decays, we plotted $10^3$ of these on top of $10^6$ above-threshold events.}. The key observation to make here is that the threshold points P(Q) in Fig.~\ref{fig:theory2}a(b) are not at all obscured when above-threshold events are superimposed; in fact, these latter serve to graphically reinforce the threshold points, located on the envelope of the collective shape.

 It should be clear, however, that without our color-coding in Fig.~\ref{fig:theory2}(a,b), it is  not possible to tell the exact position of P and Q\footnote{There is a sort of kink near P and Q, but this turns out to be related to $M_{\ell^+ \ell^-}^{max}$ only.}. Nevertheless, P and Q share a tight relationship: namely, they must uniquely identify the three endpoints $M_{4l}^{max}$, $\overline{M}_{3l}^{max}$, and
$(M_{e e}^{max})\times(M_{\mu \mu}^{max})$, this last of which is obtainable from a wedgebox plot\cite{cascades}, i.e. a plot of $M_{e e}~vs.~M_{\mu \mu}$.
We can therefore find the precise locations of P and Q with a graphical device:
superimpose a plot of $M_{4l}~vs.~\overline{M}_{3l}\frac{(M_{e e})\times(M_{\mu \mu})}{(M_{e e}^{max})\times(M_{\mu \mu}^{max})}$  with one of $M_{4l}\frac{(M_{e e})\times(M_{\mu \mu})}{(M_{e e}^{max})\times(M_{\mu \mu}^{max})}~vs.~\overline{M}_{3l}$ (i.e. superimpose Fig.~\ref{fig:theory2}a and Fig.~\ref{fig:theory2}b with swapped and rescaled axes).
This gives rise to a characteristic 'cat-eye' shape(Figure \ref{fig:theory2}c), where the upper corner of the eye pinpoints the correct threshold extrema of $M_{4l}$ and $\overline{M}_{3l}$.

There is another very useful correlation here, namely  $\overline{M}_{3l}~vs.~{M}_{4l} \overline{M}_{2l2l}$ (see previous footnote for alternatives). Analogous to the situation in Fig.~\ref{fig:theory2}a and Fig.~\ref{fig:theory2}b, threshold values of $\overline{M}_{3l}^{max}$ and  $M_{4l}^{max} \overline{M}_{2l2l}^{max}$ for this $\mathbb{X} \to {\widetilde\chi_i}^0 {\widetilde\chi_j}^0$ decay again lie on the envelope of the
 correlation shape at point `X' in Fig.~\ref{fig:theory2}d; what is more interesting, every point on the envelope \emph{above} X (say, `Y' in Fig.~\ref{fig:theory2}d) corresponds to another set of  $\overline{M}_{3l}^{max'}$ and  $M_{4l}^{max'} \overline{M}_{2l2l}^{max'}$ for a decay $\mathbb{X} \to \zeta \xi$, where $\zeta$ and $\xi$ are some hypothetical set of heavier particles ($m_{\zeta, \xi} > m_{i,j}$) which follow the same decay chain through an on-shell slepton to leptons and the LSP.
 In other words, if $\zeta$ and $\xi$ really were produced, they would have decay kinematics leading precisely to the endpoints identified at Y (we will call this a `pseudo-threshold'). Every such pseudo-threshold therefore yields a set of endpoints $\overline{M}_{3l}^{max'}$ and  $M_{4l}^{max'} \overline{M}_{2l2l}^{max'}$ which can be used to constrain $m_1$, $m_s$, and $m_\zeta$ (where without loss of generality we can take $\zeta = \xi$), this latter being merely an auxiliary parameter at each such point. What we are effectively saying here is that the shape of the envelope above X depends on $m_1$ and $m_s$ only, and
 sufficiently precise measurement of this envelope can constrain these parameters.

Now there is really no difference between this situation with a continuously-massed Higgs and that of the decay (\ref{zizjdecay}), or more specifically
\begin{equation} \label{zizjdecay2}
\mathbb{X}\to \mathbb{X}' + \widetilde{\chi}_{i}^0(\to
    \tilde{e}^\pm {e}^\mp \to e^+ e^- \widetilde{\chi}_{1}^0)~ \widetilde{\chi}_{j}^0 (\to \tilde{\mu}^\pm {\mu}^\mp
    \to  \mu^+ \mu^-  \widetilde{\chi}_{1}^0)
\end{equation}
where $\mathbb{X}$ could be $pp$, $\tilde{q} \tilde{q'}$, $\tilde{q} \tilde{g}$ ,
$\tilde{g} \tilde{g}$, $\widetilde{\chi}_{2}^\pm \widetilde{\chi}_{2}^\mp$ etc., or  any two\footnote{The present work assumes R-parity is exact, though the general technique does not require this.} SUSY particles with a continuous center of mass energy greater than or equal to $m_i + m_j$, while $\mathbb{X}'$ are any collection of particles, e.g. hadronic jets, that do not confuse the 4-lepton signal. As long as the neutralino-pair subchains are intact, the mother chain is irrelevant. Note also we do not require the lightest neutralino to be stable, as long as its decay products do not include leptons.
As we shall see in the next section this technique works very well in MC simulations with a minimal requirement of four isolated leptons plus missing energy.

\section{Examples of Applications} \label{sec:mc}
In this section we demonstrate the robustness of our technique at two different MSSM parameter points with varying decay topologies. Our MC setup uses HERWIG 6.5 and private codes as in our previous publications \cite{4l-inv,EWwedgebox,cascades} and the reader is referred to these for details.

We generate events
\begin{equation}\label{colored}
    pp \to \{\tilde{q}, \tilde{g}, \widetilde{\chi}_{1,2}^\pm, \widetilde{\chi}_{1,2,3,4}^0 \}+
    \{\tilde{q}, \tilde{g}, \widetilde{\chi}_{1,2}^\pm, \widetilde{\chi}_{1,2,3,4}^0 \}
\end{equation}
 for $10 - 30~fb^{-1}$(a low-luminosity year at the LHC) and only pass those which decay to a hard and isolated  $e^+ e^-$ and $\mu^+ \mu^-$  pair with sufficient
  missing energy($\slashchar{E_{T}} ~ > ~ 20\, \hbox{GeV}$).
  This eliminates SM backgrounds aside from $pp \to Z^*Z$, though this can be modeled and subtracted (by a Z-veto if necessary, which happens to work perfectly at the parameter points we will consider).
  SUSY  backgrounds fall into two categories: those which produce exactly four leptons and those which produce more than this (presumably losing some to isolation cuts, detector effects, etc.). The first category includes slepton or chargino `3+1' decays such as $\widetilde{\chi}_{2}^\pm (\to {l}^\pm {l'}^\pm { l'}^\mp~ \nu \nu' \overline{\nu'} ~\widetilde{\chi}_{1}^0) ~ \widetilde{\chi}_{1}^\mp (\to { l''}^\mp ~\nu'' ~\widetilde{\chi}_{1}^0)$ where one decay goes to three leptons and the other to one. These could always be substantially reduced via flavor subtraction, but this is not necessary since they have totally different kinematics from a `2+2' decay (i.e. (\ref{zizjdecay2})) and should not obscure the envelopes in plots such as Fig.~\ref{fig:theory2}.
  Moreover, such decays would exhibit their own characteristic correlation shapes in addition to the trilepton invariant mass edge. In the worst case, then, four-lepton SUSY backgrounds should be considered an `enriched signal.' The second category of SUSY backgrounds includes events such as $\tilde{\tau} \tilde{\tau}$ where each stau decays $\tilde{\tau} \to \tau \widetilde{\chi}_{2}^0 \to {l}^\pm {l'}^\pm { l'}^\mp~ \nu \nu' \widetilde{\chi}_{1}^0$ and two of the total of six leptons are somehow lost. Yet such events, if not made utterly small by leptonic branching fractions, would introduce a more-or-less diffuse halo in our correlation plots, and in particular should not confuse the identification of threshold points.

\subsection{On-Shell Box}

The most basic illustration of the HT technique for the decay
(\ref{zizjdecay2}) is when only one particular $(i,j)$-combination has a significant branching ratio; if, in addition, $i=j$, threshold endpoint formulae simplify dramatically, e.g.
\begin{equation}
{M}_{4l}^{max}= \frac{ m_j^2(2 m_j^2 m_s^2 - m_1^2  ) - m_s^4}{ m_j m_s^2}
\end{equation}
 A $\widetilde{\chi}_{2}^0 \widetilde{\chi}_{2}^0$ pair, for example,  is often the chief product of colored cascades at mSUGRA points such as SPS1a\cite{benches}. Though we have checked our technique works at SPS1a,  we choose another point which has higher rates and therefore better demonstrates the agreement with theory presented in the last section:
 \begin{description}
  \item[On-Shell Point]
   \begin{eqnarray*}
    \mu = 250\, \hbox{GeV}  ~~ ~~~~ M_2 = 250\, \hbox{GeV} ~~~~ ~~ M_1 = 125\, \hbox{GeV}\\
    \tan \beta = 10 ~~ ~~~~ M_{\tilde{\tau},(\tilde{e}, \tilde{\mu})_L } = 250\, \hbox{GeV} ~~ ~~~~ M_{(\tilde{e}, \tilde{\mu})_R} = 130\, \hbox{GeV}
      \\
       m_H = 700\, \hbox{GeV} ~~~~ M_{\tilde{q}} \approx 400\, \hbox{GeV} ~~~~ M_{\tilde{g}} \approx 500\, \hbox{GeV}
   \end{eqnarray*}
 \end{description}
 Here we have raised the left-handed slepton and stau soft mass inputs above those of the right-handed selectron and smuon (note the physical masses differ slightly from these, see Table \ref{tab:mass}) to suppress sneutrino exchange and stau modes which may reduce the magnitude of (but not character of) the signal. The heavy Higgs mass is set rather high to remove it from the analysis, but setting it lower would in fact be beneficial since Higgs contribute to the signal process $\mathbb{X} \to  \widetilde{\chi}_{i}^0 \widetilde{\chi}_{j}^0$.

\begin{figure}[!htb]
\begin{center}
\includegraphics[width=6.5in]{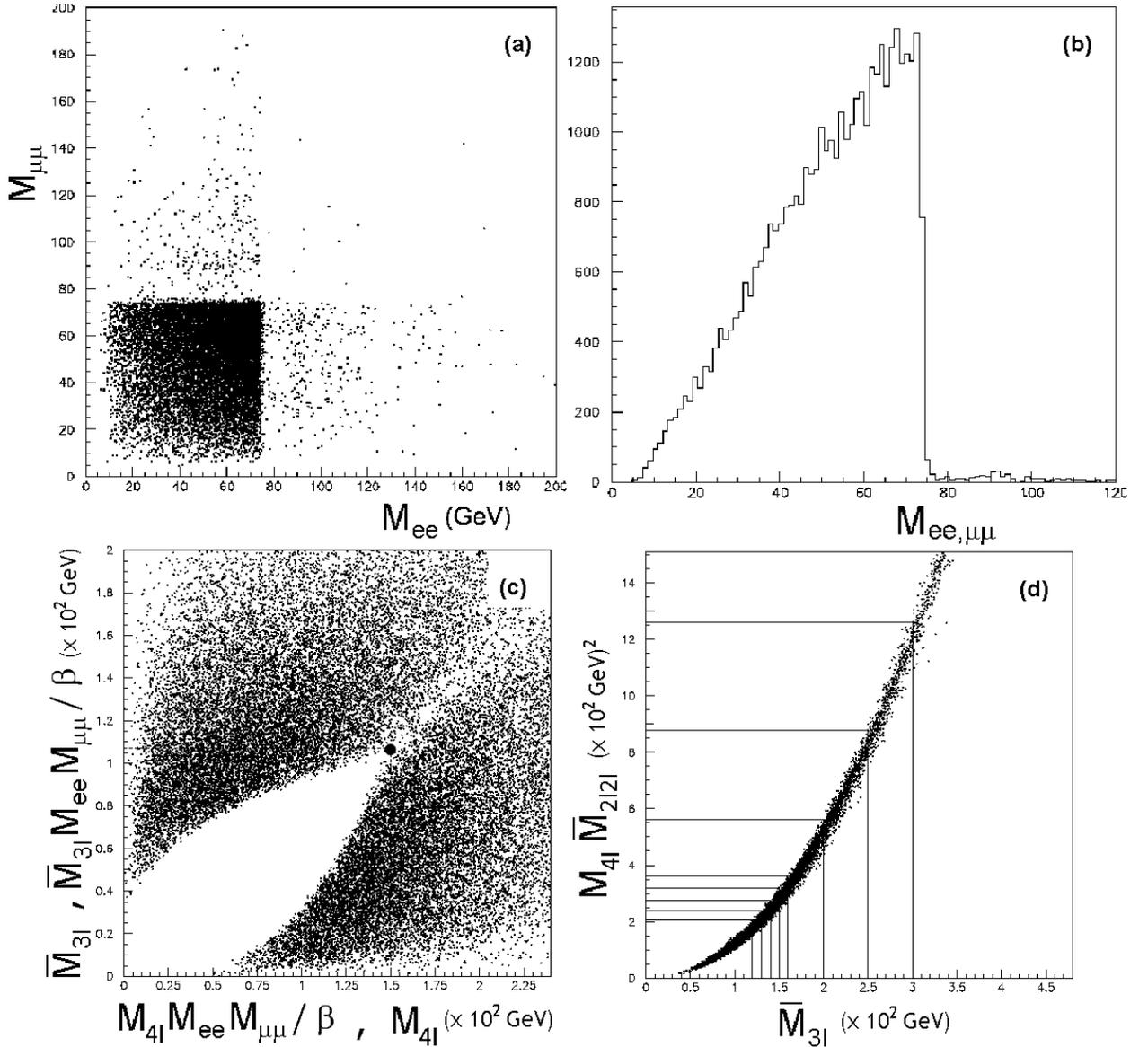}
\end{center}
\vskip -12.0cm
\caption{\small \emph{ Various plots for the On-Shell Point with $30~fb^{-1}$ luminosity. The wedgebox plot (a) is mostly box-like with a triangular dilepton mass distribution (b) typifying an on-shell decay. The threshold values of $M_{4l}$ and $\overline{M}_{3l}$ for the $\widetilde{\chi}_{2}^0 \widetilde{\chi}_{2}^0$ resonance ($\beta = (75 \, \hbox{GeV})^2$)
in the cat-eye plot (c) are easily identified at the `corner of the eye' (the black dot). We proceed to measure several points
along the envelope of the shape in (d) for $\overline{M}_{3l} > \overline{M}_{3l}^{max}$; these can be used to constrain the slepton and LSP masses.}
}
 \label{fig:onshell}
\end{figure}

\begin{table}
 \caption{\small \emph{Relevant masses  at the On-Shell Point and Off-Shell Point
 (all masses in GeV).}}
    \begin{center}
     \begin{tabular}{|c|c|c|c|c|c|c|c|} \hline
 &  ${\widetilde\chi}^0_1$
 &  ${\widetilde\chi}^0_2$
 &  ${\widetilde\chi}^0_3$
 &  ${\widetilde\chi}^0_4$
 &   $\tilde{e}_R,\tilde{\mu}_R$
 &  ${\widetilde\chi}^\pm_1$
 &  ${\widetilde\chi}^\pm_2$
 \\
 \hline
 On-Shell &  $117.1$  & $197.5$ & $257.2$ & $317.7$ & $134.7$ & $193.3$ & $317.0$  \\
 Off-Shell  &  $79.6$  & $131.5$ & $160.6$ & $249.5$ & $148.0$ & $115.6$ & $249.0$ \\ \hline
       \end{tabular}
    \end{center}
 \label{tab:mass}
\end{table}
With a luminosity of $30~fb^{-1}$  the wedgebox plot in Figure \ref{fig:onshell}a is a very dense and symmetric box structure, which strongly suggests degenerate neutralinos decaying via degenerate sleptons\footnote{Strictly speaking this could also be a chargino decaying via a sneutrino, i.e. $\widetilde{\chi}_{2}^\pm \to l^\pm \tilde{\nu} \to l^\pm l^\mp \widetilde{\chi}_{1}^\pm$. Other signs such as the number of 6-lepton events(after
$\widetilde{\chi}_{1}^\pm \to l^\pm \nu \widetilde{\chi}_{1}^0$) might resolve this ambiguity.}. Since the dilepton mass distribution(Figure \ref{fig:onshell}b) is quite triangular, we could also guess that the sleptons are on-shell ---  in the next subsection we will show how to establish this quantitatively.

Thus, being certain of having a fairly homogenous sample of  $\mathbb{X} \to  \widetilde{\chi}_{2}^0  \widetilde{\chi}_{2}^0$ decays, where each neutralino decays as  $\widetilde{\chi}_{2}^0 \to \tilde{\ell}^\pm {\ell}^\mp
    \to  \ell^+ \ell^-  \widetilde{\chi}_{2}^0$ with a dilepton edge $M_{\ell^+ \ell^-}\sim~75\, \hbox{GeV}$, we can directly proceed to construct a cat-eye plot in Fig.~\ref{fig:onshell}c to determine threshold values of
    ${M}_{4l}^{max}$ and $\overline{M}_{3l}^{max}$. Note that even at this low luminosity, the threshold point (marked as a large black dot in Fig.~\ref{fig:onshell}c) can be located to a few GeV precision: we measure
    $({M}_{4l}^{max},\overline{M}_{3l}^{max}) = (150\pm 5, 107 \pm 5)\, \hbox{GeV}$. This is quite close to the expected values of $(153.9, 112.4)\, \hbox{GeV}$ found by plugging in masses from Table \ref{tab:mass} into formulae in the Appendix. The HT method is working.

    Having determined $\overline{M}_{3l}^{max}$ from the cat-eye plot,
    we  proceed to measure a number of pseudo-threshold points in a plot of $\overline{M}_{3l}~vs.~{M}_{4l} \overline{M}_{2l2l}$ for $\overline{M}_{3l} > \overline{M}_{3l}^{max}$ (see Fig.~\ref{fig:onshell}d). Table \ref{tab:pseudo} compares some of our measurements to expected values over a liberal range of $\overline{M}_{3l}$, where the reader can verify good agreement.
     Using a large number $n$ of measurements (say $n=30$) we search $(m_1, m_s, m_{\zeta_1},m_{\zeta_2}, ... m_{\zeta_n})$-space for the best fit to the corresponding $(\overline{M}_{3l}^{max}, {M}_{4l}^{max} \overline{M}_{2l2l}^{max})_{i=1..n}$, using analytical expressions in the Appendix.
     We find $m_1 = 116 \pm 10 \, \hbox{GeV}$ with the slepton mass shifted by a near constant, $m_s = (m_1 + 17) \pm 2 \, \hbox{GeV}$. From these and the dilepton edge $M_{\ell^+ \ell^-} = ~75 \pm 1\, \hbox{GeV}$, we can likewise determine $m_2 = 210 \pm 12 \, \hbox{GeV}$.
     With a higher luminosity sample (say, $300~fb^{-1}$) error-bars could certainly be reduced by a factor of several; in this case the masses $(m_1,m_s,m_2)$ can ultimately be found to about 2\% accuracy.

\begin{table}
 \caption{\small \emph{Measurements of ${M}_{4l} \overline{M}_{2l2l}$ for various $\overline{M}_{3l}$ at the On-Shell Point (viz. Fig.~\ref{fig:onshell}d) and comparison to theoretical values (boldface) assuming these correspond to pseudo-thresholds, i.e. plugging slepton and LSP masses from Table~\ref{tab:mass} and $m_\zeta$ (in place of $m_j$) into formulae in the Appendix
 (all masses in GeV).}}
    \begin{center}
     \begin{tabular}{|c|c|c|c|} \hline
$\overline{M}_{3l}$ & ${M}_{4l} \overline{M}_{2l2l}$ & Theory & $m_\zeta$  \\
 \hline
120 & $205 \pm 1$ & \textbf{205.6} & 203.2 \\
130 & $239 \pm 1 $& \textbf{240.3}  & 211.0\\
140 & $275 \pm 2$ &\textbf{278.0} & 219.0 \\
150 & $318 \pm 2$ &  \textbf{318.0} & 227.0 \\
160 & $363 \pm 2$ & \textbf{360.9}  & 235.2\\
200 & $560 \pm 5$ & \textbf{559.9} & 269.0 \\
250 & $875  \pm 5$ & \textbf{868.0} & 313.0 \\
300 & $1260  \pm 5$ & \textbf{1250.6} & 595.5 \\
 \hline
       \end{tabular}
    \end{center}
 \label{tab:pseudo}
\end{table}

\subsection{Off-Shell Box}
\label{subsec:off}

Decays through off-shell sleptons are simpler for two reasons. First, mathematically speaking, invariant mass edges do not depend on slepton masses; in the case of a box topology these can therefore only depend on $m_1$ and  $m_j$. Since the dilepton mass edge is equal to the difference of these, $(m_j-m_1)$, we only need one other function of these masses to determine them.

Secondly, the physical degrees of freedom are easier to analyze: for threshold decays in particular, each neutralino  $\widetilde{\chi}_{j}^0$  is at rest in the center of mass frame and leptonic invariants are maximized/minimized when the same flavor leptons are emitted antiparallel/parallel. The product $(M_{e e})\times(M_{\mu \mu})$ is therefore maximal when the electron(muon) is antiparallel to the positron(anti-muon),
and is equal to  $(m_j-m_1)^2$.
 For this kinematical configuration, however, all the other invariants
$M_{4l}$, $\overline{M}_{2l2l}$, $\overline{M}_{l2l}$, $\overline{M}_{3l}$, and  $\overline{M}_{ll}$ are equal(up to a constant factor) to $(m_j-m_1)$ and therefore do not provide independent information.

The trick is to consider configurations where the electron and positron are antiparallel while the muon and anti-muon are parallel (or vice versa). In this case $M_{e e}$ or $M_{\mu \mu}$ is maximal and the other invariants (excepting $\overline{M}_{l2l}$) unconditionally attain \emph{minima} when the $e^+ e^-$ pair is emitted at right angles to the $\mu^+ \mu^-$ pair:
\begin{eqnarray} \label{offshell4l}
 \label{offshelleq}
  \overline{M}_{4l, 2l2l,3l,ll}^{min} &=& (m_j-m_1)\left(
    \frac{\alpha + \beta \frac{m_1}{m_j}  + \gamma \frac{m_1^2}{m_j^2}  }{\xi}
         \right)^{\frac{1}{4}}
\end{eqnarray}
for specific values of the parameters $\alpha$, $\beta$, $\gamma$, and $\xi$ listed in Table \ref{tab:offshell}. Thus, any one of these in conjunction with the dilepton mass edge uniquely determines $m_j$ and $m_1$. Geometrically, these endpoints are found at the intersection of the line $M_{l^+ l^-} = (m_j-m_1)$ with the near edges of the correlation shapes on a plot of  $M_{l^+ l^-}$  versus  $M_{4l}$, $\overline{M}_{2l2l}$, $\overline{M}_{3l}$, or $\overline{M}_{ll}$.  These intersections can be found with high precision and then checked against eachother for consistency.

To see the efficacy of this method,  consider the following parameter point (see Table~\ref{tab:mass} again for the physical spectrum):
 \begin{description}
  \item[Off-Shell Point]
   \begin{eqnarray*}
    \mu = 150\, \hbox{GeV}  ~~ ~~ M_2 = 200\, \hbox{GeV} ~~ ~~ M_1 = 100\, \hbox{GeV}\\
    \tan \beta = 10 ~~ ~~ M_{\tilde{\tau},(\tilde{e}, \tilde{\mu})_L } = 300\, \hbox{GeV} ~~~~ ~~~~ M_{(\tilde{e}, \tilde{\mu})_R} = 135\, \hbox{GeV}
      \\
       m_H = 700\, \hbox{GeV} ~~~ M_{\tilde{q}} \approx 300\, \hbox{GeV} ~~~ M_{\tilde{g}} \approx 350\, \hbox{GeV}
   \end{eqnarray*}
 \end{description}

\begin{figure}[!htb]
\begin{center}
\includegraphics[width=2.5in]{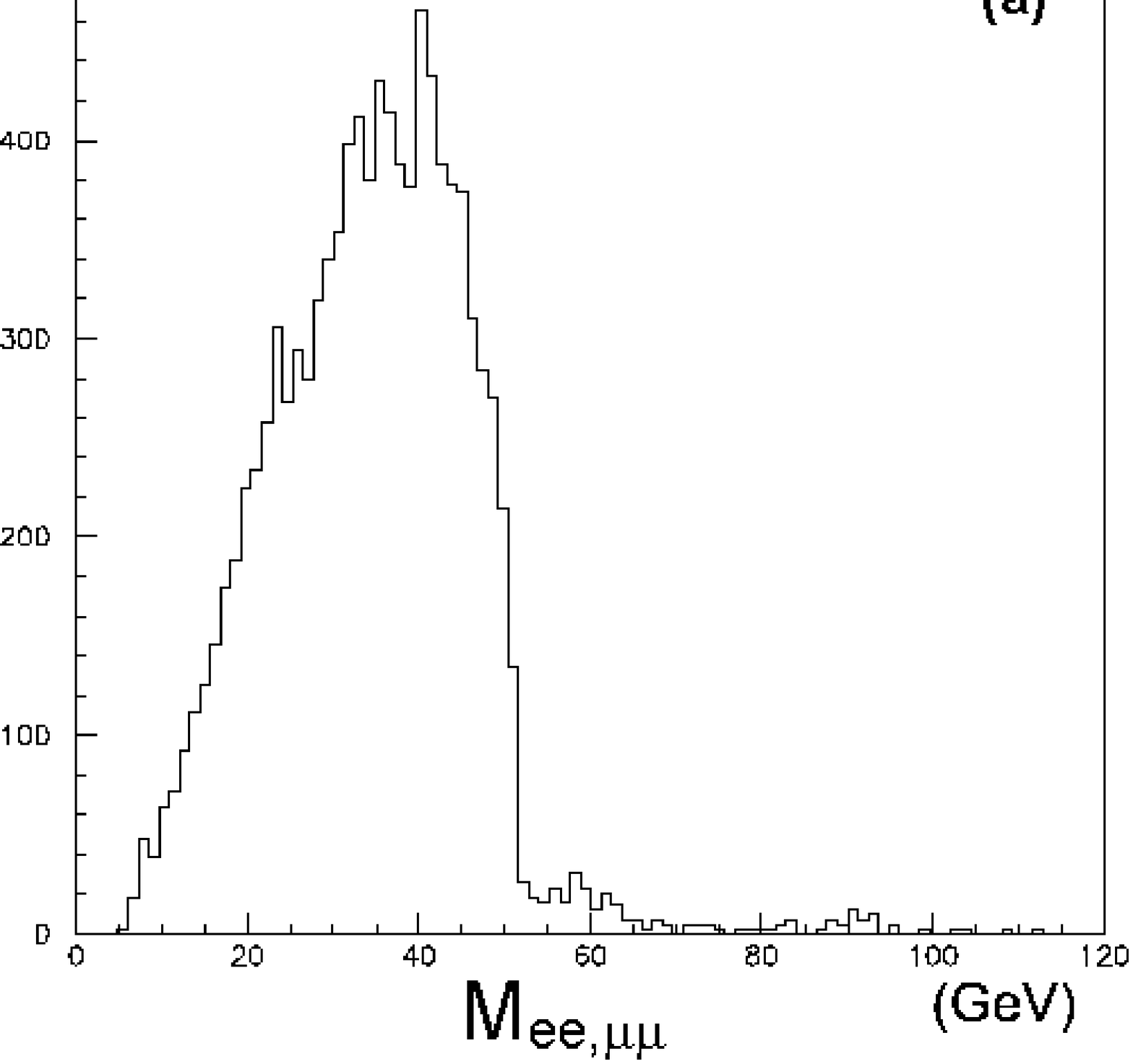}
\includegraphics[width=2.5in]{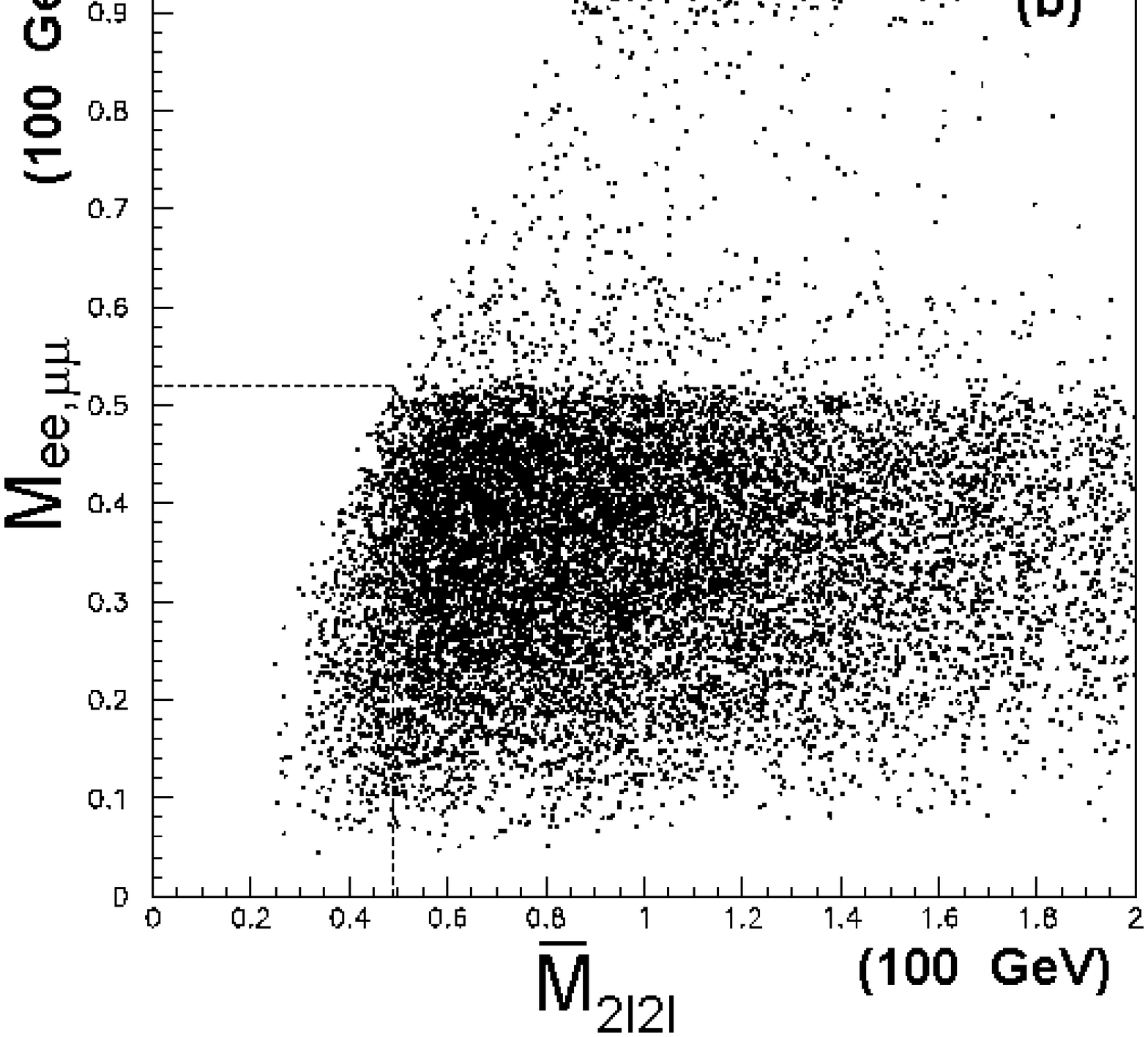}
\end{center}
\vskip -2.0cm
\caption{\small \emph{ MC Simulation at the Off-Shell Point for $10~fb^{-1}$ luminosity.  The dilepton invariant mass distribution in  (a) indicates a maximum of about $M_{l^+ l^-}^{max} \sim 52\,\hbox{GeV}$. In (b) we show where this maximum is coincident with the minimum value of $\overline{M}_{2l2l}$ at the dotted line, which is to be matched against expression (\ref{offshelleq}).
 Corresponding plots for the other invariants look very similar.}
}
 \label{fig:offshell}
\end{figure}

Though the $10~fb^{-1}$ dilepton mass distribution in Fig.~\ref{fig:offshell}a looks very similar to that of the On-Shell Point, the dilepton edge at $M_{l^+ l^-}^{max} \sim 52\,\hbox{GeV}$ is in fact due to three-body decays via off-shell sleptons. We might guess this from the vaguely less-than-triangular shape, but now we have a better way. We simply plot $M_{l^+ l^-}$ versus each of the invariants $M_{4l}$, $\overline{M}_{2l2l}$, $\overline{M}_{3l}$, and $\overline{M}_{ll}$  (e.g. Fig.~\ref{fig:offshell}(b)) and match  against the formulae in  (\ref{offshelleq}). Table \ref{tab:offshell} shows that these give mutually consistent values of $m_1$ and $m_j$, which at high luminosity may be determined to several percent or so.
Had we constructed the same plots at our On-Shell Point, we would have obtained a set of endpoints which give inconsistent, even nonsensical  values for $m_{1,j}$ (e.g. negative LSP mass!). Though the dilepton distributions of Fig.~\ref{fig:onshell}b and Fig.~\ref{fig:offshell}a give qualitative hints of whether sleptons are on- or off-shell,
here we seem to have the first definitively \emph{quantitative} method  of determining this which succeeds even with modest(several GeV) endpoint precision.

 \begin{table}
 \caption{\small \emph{Offshell Decay Parameters in (\ref{offshelleq}) and Fits to MC data at the Off-Shell Point for the LSP mass ($m_1$), assuming the dilepton edges are exact and $0.2\, \hbox{GeV}$ precision on other endpoints. Also shown are attempted fits to data at the On-Shell Point which, even with larger($\pm 1\, \hbox{GeV}$) errors,  give self-inconsistent results.}}
    \begin{center}
     \begin{tabular}{|c|c|c|c|c|c|c|} \hline
   Invariant & $\alpha$ & $\beta$ & $\gamma$ & $\xi$ & LSP(Off) &  LSP(On) \\ \hline
      $M_{4l}^{min}$ & 4 & 4 & 1 & 1 & $85 \pm 5$ & $22 \pm 5$ \\ \hline
      $\overline{M}_{2l2l}^{min}$ & 2 & 0 & 1 & 3 & $85 \pm 11$ & $172 \pm 78$ \\ \hline
      $\overline{M}_{3l}^{min}$ & 11 & 10 & 3 & 8 & $84 \pm 6$ &  $1 \pm 5$ \\ \hline
      $\overline{M}_{ll}^{min}$ & 3 & 2 & 1 & 48 & $86 \pm 15$ & $-8 \pm 10$ \\ \hline
           \end{tabular}
    \end{center}
 \label{tab:offshell}
\end{table}

\section{Conclusions} \label{sec:conc}

Let us now summarize the HT technique and how we applied it in this work:
\begin{enumerate}
  \item DEFINE the decay chain $\mathbb{X} \to \mathbb{X'} +  ABC... \to n_j ~\mathrm{jets} + n_\textit{l} ~\mathrm{leptons}$, where $\mathbb{X}$ is anything and $ \mathbb{X'}$ is exclusive of jets and leptons. \textit{In our case this was  $\mathbb{X} \to \widetilde{\chi}_{i}^0 \widetilde{\chi}_{j}^0$ where each neutralino then decayed via a slepton to a pair of leptons and the LSP.}
  \item DERIVE analytic expressions for all jet and lepton invariant mass endpoints as functions of NP masses, and find kinematic configurations for which two or more of these are extremal at threshold production. \textit{In our case of four leptons($n_l=4$ ), there are in principle 7 independent invariants; of these, we found $(M_{e e})\times(M_{\mu \mu})$,
       $M_{4l}$, $\overline{M}_{2l2l}$, $\overline{M}_{l2l}$, $\overline{M}_{3l}$, and  $\overline{M}_{ll}$
       were simultaneously maximal for on-shell decays, wheareas for off-shell decays $M_{e e}$($M_{\mu \mu})$ was maximal(minimal) where $M_{4l}$, $\overline{M}_{2l2l}$,  $\overline{M}_{3l}$, and  $\overline{M}_{ll}$ were minimal.}

  \item  DISPLAY correlations of the above invariants in a scatterplot which  makes their threshold endpoints visually obvious. Use these to solve for NP masses.\textit{ For off-shell slepton decays it sufficed to plot  each invariant versus $M_{l^+ l^-}$ and note the intersection with the line $M_{l^+ l^-}=M_{l^+ l^-}^{max}$. For on-shell decays we found it useful to make a cat-eye plot to find the threshold value of $\overline{M}_{3l}^{max}$, followed by measuring
      several pseudo-thresholds in a plot of $\overline{M}_{3l} ~ vs.~ M_{4l} \overline{M}_{2l2l}$. Matching to analytical expressions constrains $(m_1, m_2, m_s)$ to within a few percent of their nominal values.}
\end{enumerate}

Let us add a few remarks on these three steps. The first step depends of course on the specific NP model under consideration, but once a decay chain has been selected,
deriving analytical formulae for endpoints in the second step is not difficult: the sought-after kinematic configuration is usually where each jet or lepton is emitted at a polar angle of $\theta = 0$ or $\theta = \pi$  or perhaps $\theta = \pi/2$ (for minima)  in the frame of the decaying parent particle, backwards Lorentz-boosting to the center-of-mass frame (see Appendix of \cite{4l-inv} for an example).
Choosing which invariants to plot in the third step is a matter of trial-and-error ---  kinematic simulation (as in Fig.~\ref{fig:theory1} and  Fig.~\ref{fig:theory2}) can be used to see which correlations have a visible threshold point when above-threshold decays are superimposed.

To demonstrate the general applicability of this technique to NP, we present below a  partial spectrum of examples. Each example in itself entails numerous variations(e.g. on-shell intermediate states could likewise be taken off-shell):
\begin{itemize}
 \item Neutralino Decay via higgs:
    $\widetilde{\chi}_{2}^0 \widetilde{\chi}_{2}^0(\to \widetilde{\chi}_{1}^0 h(\to bb))$.
   This $n_j = 4$ decay is a potential competitor to the sleptonic modes considered in this paper.
 \item Sneutrino Pair Production:
   $ \tilde{\nu} \tilde{\nu}(\to l^\pm \widetilde{\chi}_{1}^\mp (\to  \widetilde{\chi}_{1}^0 W^-(\to l'^\mp \nu' )))$
  Here $n_l = 4$ as for neutralino pair production, but with totally different kinematics: one can analogously define $M_{4l}$, $\overline{M}_{2l2l}$, $\overline{M}_{l3l}$, etc., though threshold extrema of these will have a different correlation.
  \item Chargino Pair Production:
    $\widetilde{\chi}_{2}^\pm \widetilde{\chi}_{2}^\mp(\to l^\mp \tilde{\nu}(\to l^\pm \widetilde{\chi}_{1}^\mp (\to l^\mp \nu {\widetilde{\chi}_1}^0)))
  $
     This is a $n_l = 6$ case which means there are as many as 45 relativistic invariants\footnote{The number of invariants is computed as the number of pairwise contractions among the observable momenta, as well as with powers of $\epsilon^{\mu \nu \rho \sigma}$}.
    \item Slepton Pair Production
  $\tilde{l}^\pm \tilde{l}^\mp (\to l^\mp \widetilde{\chi}_{2}^0(\to l^\pm l^\mp \widetilde{\chi}_{1}^0))$. This could either be $n_l = 6$  or $n_l = 4$ depending on how the other slepton decays.
   \item Squark Pair Production
  $\tilde{q} \tilde{q} (\to q \widetilde{\chi}_{2}^0 (\to l^\pm \tilde{l}^\mp (\to  l^\mp \widetilde{\chi}_{1}^0)))$. A $n_j=2$, $n_l=4$ decay with 45 invariants to analyze.
   \item Gluino Pair Production
  $\tilde{g} \tilde{g} (\to q \tilde{q}(\to q \widetilde{\chi}_{2}^0 (\to q'q' \widetilde{\chi}_{1}^0))$
  This $n_j = 8$ decay has an astounding $588$ invariants.
  \item Nonstandard Higgs decays $h \to X Y$(e.g. \cite{nonstandh})
  \item Exotic Vectors (from, e.g. Little Higgs \cite{littlehiggs})
  \item Kaluza-Klein Pair Production (e.g. \cite{kk}).
   \item Exotica decays to top-pairs $pp\to X \to t \overline{t}$ (e.g. \cite{ttbar})
    \item Low-Scale Technicolor  (e.g. $W^\pm \pi_T \to l^\pm \nu b \overline{b}$ \cite{techni}).
\end{itemize}

The hidden threshold method provides another way to see signatures of these NP models and determine any unknown masses involved. Results can be made even stronger in combination with inclusive techniques and complementary methods of mass determination (in SUSY for example, see \cite{atlas, massdet, massrel1}).
We hope that research along these lines will allow an earlier discovery of NP at hadron colliders.

\section*{Appendix}

Threshold maxima for on-shell decays are readily computed via the methodology
of \cite{4l-inv}. For the case $i=j$ we just substitute $m_A = 2 m_j$ in the formulae of that work for the $[++--]$ configurations, obtaining the following:
{\small
\\
\\
\noindent\(
\mathbf{{M}_{4l}^{max}}= \frac{ m_j^2(2 m_j^2 m_s^2 - m_1^2  ) - m_s^4}{ m_j m_s^2}\)
\\
\\
\noindent\(
\mathbf{\overline{M}_{2l2l}^{max}}=
\frac{1}{3^{1/4}
  m_j m_s^2} (3 m_1^8 m_j^8 - 4 m_1^6 m_j^6 m_s^2 (2 m_j^2 + m_s^2) -
   4 m_1^2 m_j^2 m_s^6 (8 m_j^6 - 12 m_j^4 m_s^2 + 6 m_j^2 m_s^4 + m_s^6) +
   m_s^8 (16 m_j^8 - 32 m_j^6 m_s^2 + 24 m_j^4 m_s^4 - 8 m_j^2 m_s^6 +
      3 m_s^8) + 6 m_1^4 (4 m_j^8 m_s^4 - 4 m_j^6 m_s^6 + 3 m_j^4 m_s^8))^{
 1/4}\)
\\
\\
\noindent\(
\mathbf{\overline{M}_{l2l}^{max}}= \frac{1}{6^{1/4}
   m_j m_s^2}(3 m_1^8 m_j^8 - 2 m_1^6 m_j^6 m_s^2 (5 m_j^2 + m_s^2) +
   6 m_1^4 m_j^4 m_s^4 (3 m_j^4 - m_j^2 m_s^2 + m_s^4) -
   2 m_1^2 m_j^2 m_s^6 (8 m_j^6 - 6 m_j^4 m_s^2 + 3 m_j^2 m_s^4 + m_s^6) +
   m_s^8 (8 m_j^8 - 16 m_j^6 m_s^2 + 18 m_j^4 m_s^4 - 10 m_j^2 m_s^6 +
      3 m_s^8))^{1/4}
   \)
\\
\\
\noindent\(
\mathbf{\overline{M}_{3l}^{max}}= \frac{\left( m_1^4 m_j^4 - 2 m_1^2 m_j^4 m_s^2 +
    2 m_j^4 m_s^4 - 2 m_j^2 m_s^6 + m_s^8\right)^{1/4}
\sqrt{m_1^2 m_j^2 - 2 m_j^2 m_s^2 + m_s^4}}{2^{1/4} m_j m_s^2}\)
\\
\\
\noindent\(
\mathbf{\overline{M}_{ll}^{max}}= \frac{\sqrt{m_1^4 m_j^4 - 2 m_1^2 m_j^4 m_s^2 + 2 m_j^4 m_s^4 - 2 m_j^2 m_s^6 +
   m_s^8}}{24^{1/4} m_j m_s^2}\)
\\
\\

 }

\end{document}